**Differentiable Land Model Reveals Global Environmental Controls on Latent Ecological Functions**

Jianing Fang[1*], Kevin Bowman[2], Wenli Zhao[1], Xu Lian[1], and Pierre Gentine[1*]

**Affiliations**

1. Department of Earth and Environmental Engineering, Columbia University, New York, NY, USA
2. Jet Propulsion Laboratory, California Institute of Technology, Pasadena, CA, USA

Corresponding authors: Jianing Fang (jf3423@columbia.edu), Pierre Gentine (pg2328@columbia.edu)

## Abstract

Do ecosystems primarily reflect evolutionary history or current environment? Predicting land–atmosphere exchange hinges on this unresolved question. Plant traits adapt to particular environments over evolutionary timescales, yet their individual relationships with current climate and soils are often obscured by limited sampling, plant-type effects, and multiple adaptive strategies that can yield similar outcomes. Crucially, it is the *coordination of traits*, rather than any single trait, that governs vegetation dynamics and ecosystem fluxes, yet such multivariate relationships cannot be directly observed. We present DifferLand, a differentiable hybrid model that integrates process understanding with machine learning to uncover latent trait–environment relationships from global satellite and *in-situ* observations (2001–2023). DifferLand explains up to 88% of the variance in canopy structure, photosynthesis, and carbon exchange by learning latent ecological axes—leaf economics, plant stature, and cropland distribution—that link long-term adaptation with short-term dynamics. Interpretable machine learning shows that these coordinated axes emerge from nonlinear interactions between plant-type attributes and local environment. Embedding such relationships into terrestrial models establishes a pathway toward adaptive models that better predict ecosystem resilience under climate change.

## Introduction

Understanding how environmental gradients shape the spatial distribution of vegetation functions is a long-standing question in ecological and climate research. Modern ecology theory posits that abiotic environmental filters[1] constrain viable plant functional trait combinations within bioclimatic envelopes, while local biotic interactions, dispersal limits, and disturbance histories further shape current plant trait distributions[2]. These interacting eco-evolutionary processes underpin an ongoing debate (Fig. S1): are spatial variations in plant traits and the ecological functions they provide primarily explained by universal scaling relationships with abiotic gradients (i.e., the functional convergence hypothesis)[3], or do species or plant-type–specific controls dominate[4], limiting the ability of environmental gradients[4] to predict within–plant-type trait variation and their roles in the carbon cycle?

If climate and other environmental gradients constrain the set of viable functional traits, it follows that they should strongly predict the current distribution of physiological and morphological plant traits. However, evaluations of univariate trait-environment relationships found environmental variables, such mean temperature, water availability, and soil properties, typically each explained less than 10-20% of the trait variations[4]. In contrast, global multivariate analyses of trait–trait relationships reveal that plant functional traits covary along key axes[5–7]—such as the leaf economics spectrum[8] and the allometry continuum[9], reflecting the coexistence of multiple adaptive strategies shaped by trade-offs under natural selection. These contrasting patterns suggest that while abiotic gradients influence trait distributions, their effects are often expressed through integrated trait combinations shaped by multiple constraints, rather than through universal relationships between individual traits and single environmental predictors[4].

Capturing the complex interplay between plant traits and environmental drivers requires moving beyond fixed plant types or univariate trait–environment relationships[10], toward models that represent how multiple traits jointly shape vegetation dynamics through biological interactions and biome-specific influences. However, current terrestrial biosphere models (TBMs) typically prescribe fixed parameter sets for each plant functional type (PFT), assuming uniform trait distributions within broad life-form categories (e.g., deciduous vs. evergreen, broadleaf vs. needleleaf)[11]. This assumption has long been criticized for neglecting local adaptation and acclimation to microclimate, topography, and disturbance, as within-PFT variation can be as large as differences among PFTs[12–14]. Directly specifying multiple spatially explicit trait-based parameters is also impractical due to the high dimensionality of trait diversity and the challenge of scaling from species to ecosystem levels[15,16].

Nonetheless, despite these limitations, mechanistic TBMs still encode key physiological processes—photosynthesis, respiration, allocation, turnover, and responses to environmental stress—and thus provide our best process-based approximation of how traits, represented as model parameters, together mediate carbon and water exchanges with the atmosphere[4]. This perspective motivates our central hypothesis: that multivariate trait–environment relationships may be learned by inverting a TBMs using observed global vegetation dynamics and spatial environmental predictors of plant traits. The tradeoffs and covariation among high-dimensional ecological parameters, together with the nonlinear interactions of environmental gradients, can be effectively represented in an 'ecological latent space'[17]—a physics-informed machine learning–derived low-dimensional embedding[18,19] of global ecological functions—that enhances predictions of land–atmosphere carbon and water exchange beyond models relying on PFT-based parameterizations.

In this study, we introduce DifferLand, a differentiable terrestrial biosphere model that learns global trait–environment relationships directly from satellite and *in-situ* observations. These relationships allow the model to capture both the long-term adaptation of vegetation to prevailing environmental conditions and its short-term sensitivity to seasonal and interannual meteorological variability, yielding more accurate predictions of global vegetation dynamics and carbon exchange than models that rely solely on PFT classifications or sparse trait–environment linkages. We find that the retrieved trait-environment relationships arise from non-linear interactions between plant-type–specific attributes and local environmental gradients. Moreover, the organization of traits along major ecological axes emerges naturally from observed vegetation dynamics, providing a key source of predictability for spatial variations in ecosystem functioning within the high-dimensional space of plant functional diversity.

## Results

### Learning Trait-Environment Relationships via Differentiable Modeling

DifferLand is a fully differentiable hybrid terrestrial biosphere model that unifies neural-network learning of global trait–environment relationships with process-based simulation of local carbon–water dynamics (Fig. 1). A global spatialization network (Fig. 1b) infers latent ecological parameters (Fig. 1c) from environmental predictors (Fig. 1a), which are then passed to a mechanistic model (Fig. 1d) that resolves monthly carbon uptake, respiration, fire carbon emissions, and associated changes in land carbon pools. Differentiability enables end-to-end optimization of these trait–environment relationships by propagating observation–model mismatches (Fig. 1e) back through the mechanistic model at each timestep.

By coupling large-scale environmental controls with grid-cell-level process realism, DifferLand can reveal potential functional relationships linking plant traits to climate, soils, and vegetation history, while retaining sensitivity to local meteorological forcing. Its flexible data assimilation framework integrates diverse constraints (Fig. S5)—from satellite observations and eddy covariance fluxes to global soil carbon and atmospheric inversions—allowing the learned relationships to capitalize on spatial correlations, trait covariation, and the physical consistency of process-based modeling. This approach provides a unified, observation-constrained framework for uncovering the environmental determinants of vegetation function and improving predictions of the terrestrial carbon cycle.

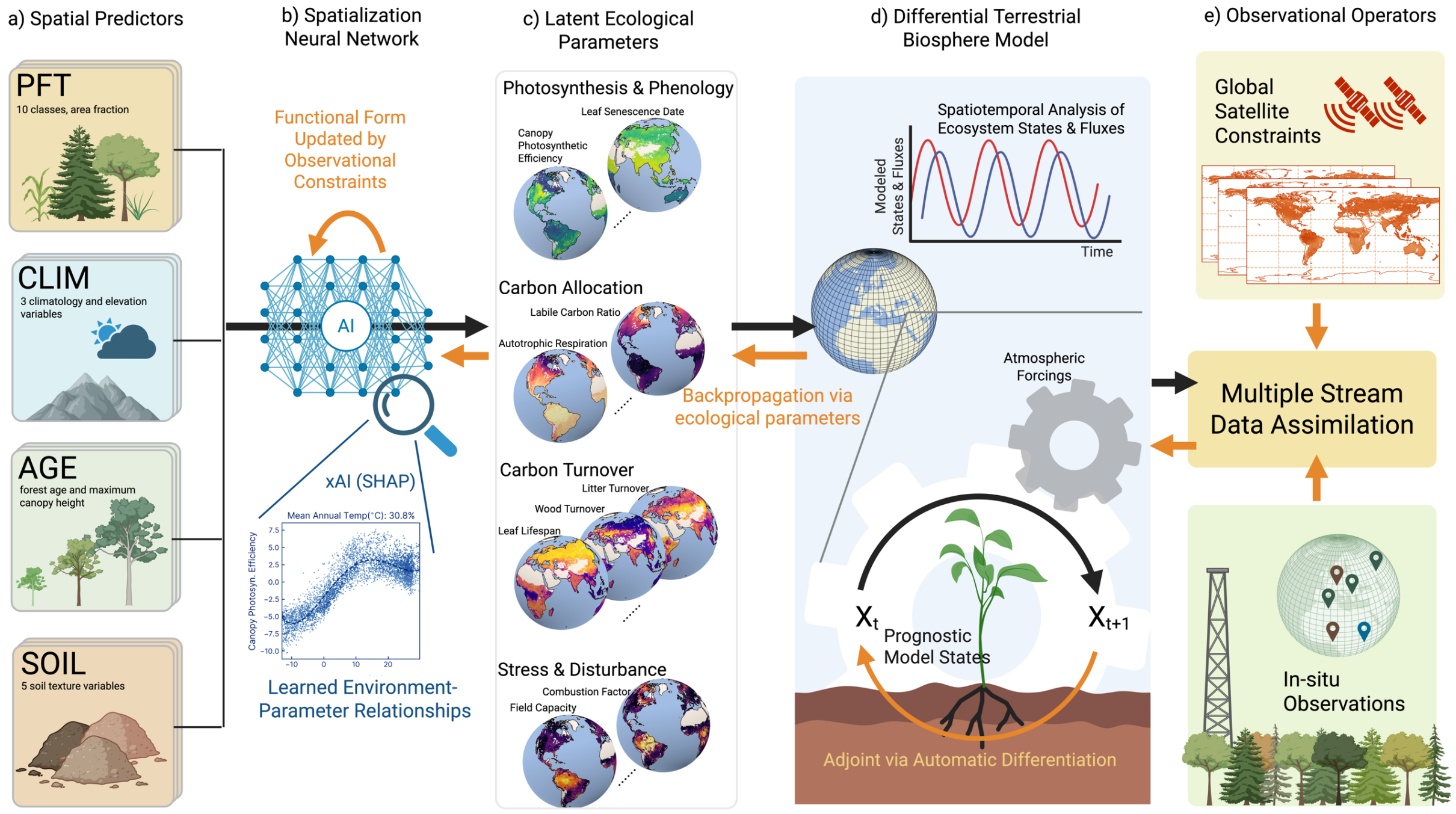

**Fig. 1** | Schematics of the DifferLand framework. a) DifferLand integrates spatial information in four groups of environmental predictors, including plant functional types (PFTs), climatology and elevation (CLIM), forest age and maximum height (AGE), and soil texture (SOIL). b) During a forward pass (black arrows), a spatialization neural network models trait–environment relationships, mapping these predictors to c) global fields of 40 ecological parameters representing plant functional traits such as photosynthesis, carbon allocation, turnover rates, and sensitivity to drought and fire. d) A terrestrial biosphere model uses these parameters to simulate ecosystem states and fluxes under meteorological forcing. e) DifferLand evaluates the simulated vegetation dynamics against multiple streams of satellite and in situ-observations, and it computes the sensitivity of simulated vegetation dynamics to trait-based ecological parameters through its differentiable structure (orange arrows). The gradient information is used to update the parameters within the spatialization neural network and iteratively optimize the learned trait-environment relationships.

### Environmental Control of Ecological Functional Parameters

To evaluate whether trait–environment relationships provide independent information for predicting global ecological dynamics beyond what is captured by PFT-based categories, we trained models using either only PFT fractions as spatial predictors or a combination of PFT fractions and environmental predictors reflecting climatology, forest age and growth potential, and soil properties. Note that in both cases, all tunable ecological parameters in the process-based model, as well as the initial values of the carbon and water pools, are optimized against observations using the same meteorological forcings via the spatialization network. Thus, the performance differences reflect the information content of the spatial predictors rather than differences in default parameter choices, model initialization, or structural assumptions.

We train and evaluate the model on satellite-based indices that include leaf area index (LAI)[20] as a measure of canopy structure, a solar-induced fluorescence (SIF) based photosynthesis proxy[21], a top-down inversion of net biosphere exchange (NBE) constrained by column integrated $CO_2$ concentration from satellites[22], satellite gravimetry-based anomalies of total equivalent water thickness (EWT) over land[23], and a global evapotranspiration (ET) product derived from a satellite data-constrained model[24]. We performed detailed sensitivity analyses and found the results to be robust to the choices of alternative datasets (Text S4).

Experiment results demonstrate that environmental gradients provide essential spatial information beyond plant functional types (PFTs) for predicting vegetation trait distributions. When incorporating all environmental predictors (PFTs, climate variables, forest age, and soil properties), DifferLand effectively captures the spatial and temporal patterns of both *in-situ* and remotely sensed observations of vegetation dynamics from 2001 to 2023 (Fig. S15 & Fig. S16). It generalizes well to held-out pixels, achieving a total spatiotemporal $R^2$ values of 0.88± 0.01 for LAI, 0.76± 0.01 for SIF, 0.71± 0.03 for NBE, 0.68± 0.02 for ET, and 0.45± 0.01 for EWT. The model also accurately captures the mean global biomass (Fig. S22a) and reasonably reproduces the assimilated trends in biomass over the past two decades (Fig. S22b). Notably, the model demonstrates minor differences in predictive performance between the training and test pixels, suggesting the model generalized well at unseen pixels (Fig. S15 & Fig. S16). In contrast, the baseline model relying solely on PFT fractions yields substantially lower total $R^2$ scores of 0.86 ±0.01, 0.73± 0.01, 0.60±0.03, 0.52±0.02, and 0.42±0.01 for LAI, SIF, NBE, and ET, and EWT respectively, when evaluated over held-out pixels, and exhibits a total spatiotemporal mean absolute error (MAE) that are 3–20% higher than those of the full environmental predictor configuration (Fig. 2a). These findings indicate PFTs alone do not fully capture the spatial variability in ecological parameters required to explain the observed global vegetation dynamics recorded by satellite and atmospheric inversion data.

However, the results also indicate that environmental variables alone are insufficient to capture spatial variations in latent ecological traits represented by the model parameters. While using only climatology, soil, and age predictors reduces MAE for ET predictions by 15% compared to the PFT-only baseline, plant-type-specific information remains crucial for explaining variability in canopy structure, as reflected in LAI, and in photosynthetic activity, as measured by SIF (Fig. 2a). When both PFT fractions and environmental variables are used as predictors, we observe larger reduction in errors compared to using either PFTs or environmental variables alone (Fig. 2a). For EWT, changes are relatively minor regardless of the choice of spatial predictors, likely because total water storage anomalies at the coarse spatial scale ($4° \times 5°$) are primarily driven by meteorological anomalies rather than spatial variations in ecological parameters. Overall, these findings suggest that interactions between PFTs and environmental variables are key to explaining variations in ecological functions related to vegetation growth and the carbon cycle, while water cycle dynamics are more strongly governed by environmental conditions or meteorological forcings.

To evaluate the independent contributions of different groups of spatial predictors, we performed a full factorial experiment in which we systematically included or excluded each group of predictors—PFT, CLIM, AGE, and SOIL—and assessed their impact on the model's ability to capture global ecological dynamics. Because of potential confounding signals among these predictor groups, we applied a hierarchical partitioning algorithm (see Methods) to disentangle their unique contributions. This method accounts for the interaction effects across different predictor groups and ensures that the sum of independent effects attributed to each predictor group equals the model's performance when using the full set of predictors. With these metrics, we mapped the dominant predictor group across regions to identify which spatial predictors most strongly explain ecological dynamics within different biomes (Fig. 2b-f).

Hierarchical partitioning reveals that plant functional type (PFT) distribution plays a dominant role in regulating carbon cycle dynamics, emerging as the most important predictor for LAI (Fig. 2b), NBE (Fig. 2c), and SIF (Fig. 2d) across 41–43% of land pixels. Spatially, the explanatory power of PFTs is positively correlated with local land cover heterogeneity (Spearman's $r = 0.32$, $p < 0.001$), particularly in regions with sharp ecotones or steep land-use intensity gradients—such as the periphery of the Amazon Basin and the Sahel—where PFT fractions explain the largest variance. In contrast, the relative importance of PFTs is lower for water cycle variables such as EWT and ET, dominating the explanation in only 27–30% of the pixels. Environmental predictors, on the other hand, provide crucial information for regional variations in ecological parameters within each PFT (Fig. S24 & Fig. S25). Specifically, CLIM variables exert greater influence in mid-latitude herbaceous biomes characterized by strong climatological gradients, such as the Eurasian Steppe and the North American Great Plains. AGE variables (e.g., estimated forest age and maximum canopy height) are important for perennial vegetation and help differentiate vegetation function in mixed tree–grass systems. While SOIL variables play a minor role in explaining carbon cycle dynamics, they are essential for predicting EWT and ET, particularly in sparsely vegetated regions (Fig. 2d–f).

To understand the implications of parameter spatialization for simulated carbon dynamics, we evaluated DifferLand's simulated carbon and water fluxes against eddy covariance observations and an ensemble of state-of-the-art land surface models. Comparison of model prediction of gross primary productivity (GPP), ET, and ecosystem respiration (RECO) against 180 eddy covariance sites (Table S4 and Fig. S33) suggests the model achieved good agreement with site-level fluxes in mean spatial gradients (Fig. S21), achieving spatial correlation of 0.87 for GPP, 0.81 for RECO, and 0.72 for ET across all sites (spatial correlation of 0.78, 0.85, and 0.87 on held-out sites) and effectively captured temporal variations of GPP, ET, and RECO across the eddy covariance sites (Fig. S19 & Fig. S20). These results are consistent across different filtering thresholds to reduce the spatial mismatch between eddy covariance tower and the model grid cells (Text S5). Furthermore, DifferLand closely reproduced the seasonal cycle (Fig. S12) annual anomalies (Fig. S11), and decadal trajectories (Fig. S10) of the assimilated CMS-Flux net biome exchange (NBE) dataset derived from atmospheric inversions for 2010–2022. The model also showed robust performance across different atmospheric inversion products and was able to largely reproduce global interannual variability in carbon fluxes during periods preceding the availability of satellite observations of column-integrated $CO_2$ (Text S5). Despite its comparatively simplified process representation, DifferLand achieved significantly lower root mean squared errors and comparable correlation with the atmospheric inversion dataset than the much more structurally sophisticated dynamic global vegetation models (DGVMs) in the TRENDYv12 S3 ensemble[25] (Table S6), both globally and regionally (Fig. S8-12). These results suggest that uncertainties in model parameters are a major source of error in simulating land–atmosphere carbon fluxes at decadal timescales, and that hybrid modelling of trait–environment relationships offer a promising avenue to reducing these uncertainties.

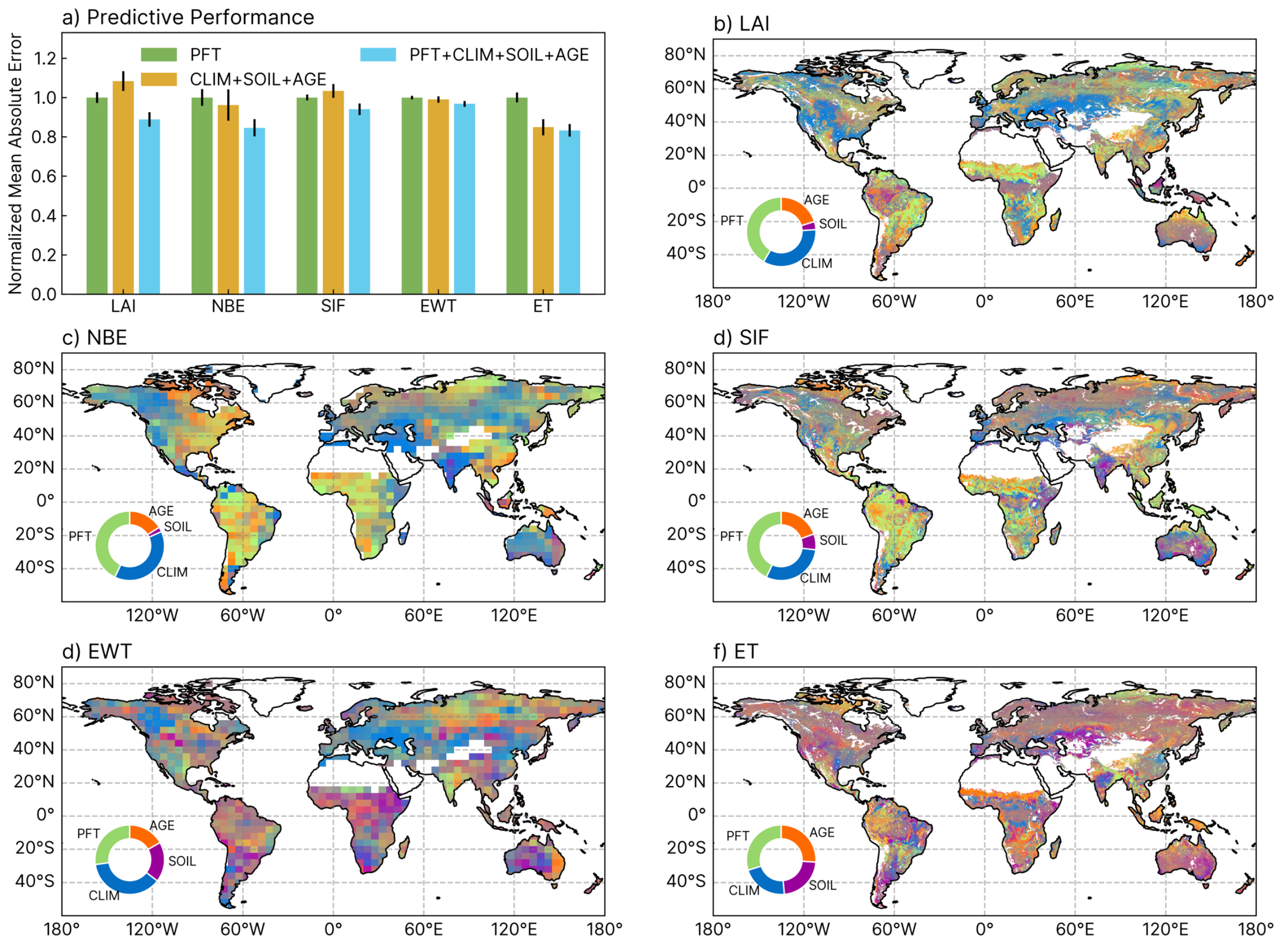

**Fig. 2** | Hierarchical partitioning of explained variance by four predictors groups. Panel a) compares the total spatiotemporal mean absolute error on held-out pixels for LAI, NBE, SIF, EWT, and ET across three model configurations between 2003 and 2023: PFT-only, environmental variables only (CLIM+SOIL+AGE), and combined PFT and environmental variables (PFT+CLIM+SOIL+AGE). Errors are normalized relative to the PFT-only configuration. Panels b-f) show the proportion of temporal variance explained in LAI, NBE, SIF, EWT, and ET within each grid cell during the same period, attributed to the model using ecological parameters predicted from spatial information in PFT, CLIM, AGE, and SOIL variables, respectively. The inset shows the proportion of pixels where each predictor group (PFT or environmental variables) has the dominant effect.

## Spatial Coordination of Ecological Parameters

DifferLand's ability to robustly predict vegetation dynamics across space by leveraging environment–parameter relationships suggests that global vegetation patterns may give rise to globally convergent plant functional traits, which in turn enable spatial predictability of the terrestrial carbon cycle. To investigate this hypothesis, we conducted a principal component analysis (PCA) on 13 ecologically meaningful and spatially identifiable parameters (see "Parameter Identifiability Analysis" in Methods and Text S6) retrieved from the model using the full set of spatial predictors to identify covarying traits and the primary axes of spatial variability (Fig. 3a). To further examine how environmental drivers influence the covariation among ecological parameters, we applied Shapley Additive exPlanations (SHAP) to the ecological axes obtained by projecting the predicted parameters onto the principal components (PCs) using the loadings from the PCA (Fig. 3e-g). We assessed the overall importance of each spatial predictor by calculating its mean absolute SHAP value, and quantified the strength and directionality of its influence using the difference in SHAP values between the upper (Q3) and lower (Q1) quantiles. This approach retains the non-linear trait–environment relationships captured by the spatialization network while simplifying the analysis of parameter covariation through a linear transformation of the ecological space.

The first principal component (PC1), explaining 53.2% of the variance, reflects the well-established leaf economics spectrum, distinguishing between plants that invest in long-lived leaves with high leaf carbon mass per area (LCMA) and those with shorter-lived leaves characterized by lower carbon investment and greater allocation to labile carbohydrates supporting seasonal leaf onset (Fig. 3b). The second component (PC2), explaining 22.1% of the variance (Fig. 3a,c), captures the tall–short vegetation gradient (Fig. 3g): short, often herbaceous vegetation exhibits high photosynthetic capacity and rapid litter turnover, whereas taller vegetation tends to have lower photosynthetic efficiency and allocates more carbon to woody biomass. These two ecological axes are well documented in trait databases at the species and community levels[5,6,9], but here we show they also emerge from satellite-observed vegetation dynamics without imposing explicit trait–environment relationships. Interestingly, a third component (PC3), explaining 10.5% of the variance (Fig. 3b), highlights regions with high photosynthetic efficiency and intensive cropland use (e.g., the U.S. Midwest, Southern Europe, and Eastern China), potentially reflecting the artificial selection of high-yield cultivars or the prevalence of photosynthetically efficient C4 crops such as maize.

Mean annual temperature emerged as the primary predictor for all three principal components (PCs; Fig. 3e–g), reflecting a first-order energy control on plant functional traits. In warmer climates, plants tend to exhibit traits that support longer growing seasons, higher photosynthetic efficiency, greater allocation to woody biomass, and taller canopy height—strategies advantageous for light competition and perennial growth. In contrast, plants in colder environments adopt traits such as shorter stature, reduced leaf lifespan, and increased carbon allocation to belowground biomass. Maximum canopy height, a proxy for growth potential, is most strongly associated with PC2. Despite the dominant role of climate, plant-type predictors also play a key role in shaping the ecological axes. Although a model configuration using only environmental predictors (CLIM+SOIL+AGE) can broadly reproduce the spatial patterns of PC1 and PC2 (Fig. S27), the inclusion of plant functional type (PFT) information—particularly cropland extent—is essential to capture the spatial variability related to managed crop productivity represented in PC3. In contrast, model configurations using only PFT-based predictors are not able to capture the same set of coordinated axes (Fig. S28). These findings suggest that while the covariation of ecological parameters primarily reflects macroecological environmental gradients, it also bears the imprint of present-day plant type distributions and land-use legacies.

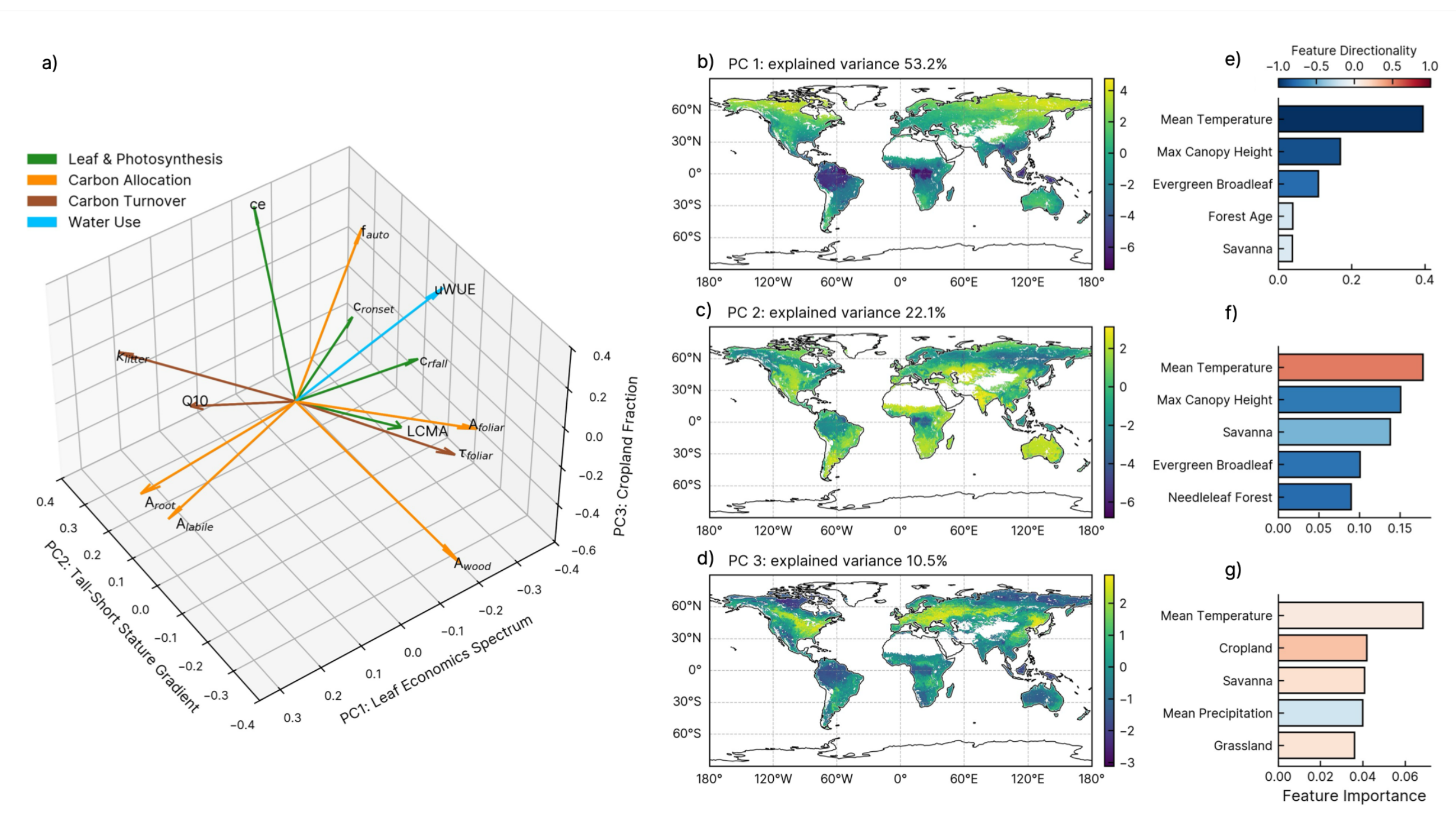

**Fig. 3 | Coordination of latent ecological functional traits along three principal axes.** a) Principal component analysis (PCA) of 13 latent ecological parameters inferred from the ensemble mean of the PFT+CLIM+SOIL+AGE model configuration. The loadings that project each parameter onto the first three PCs are plotted in the 3D plot. The 13 parameters include canopy photosynthetic efficiency (ce), underlying water use efficiency (uWUE), temperature sensitivity of heterotrophic respiration (Q10), autotrophic respiration fraction ($f_{auto}$), litter turnover ($k_{litter}$), leaf carbon mass per area (LCMA), leaf onset ($c_{ronset}$) and fall duration ($c_{rfall}$), leaf lifespan ($\tau_{foliar}$), and allocation fraction of GPP to labile ($f_{labile}$), foliar ($A_{foliar}$), wood ($f_{wood}$), and fine roots ($f_{root}$) pools; b-d) plot the spatial maps of the top three PC scores. Parameters are grouped into four categories: leaf & photosynthesis (green), carbon allocation (orange), carbon turnover (brown), and water use (blue). The three principal axes are interpreted as: PC1 – *Leaf Economics Spectrum*, PC2 – *Tall–Short Stature Gradient*, and PC3 – *Cropland Fraction*. b–d) Spatial distribution of PC1–PC3 scores projected from the spatialization neural network onto the PCA axes, with the proportion of variance explained by each component labeled. e–g) SHAP-based feature attribution for each principal component, showing the influence of spatial predictors. Bars represent mean absolute SHAP values (feature importance), while colors indicate the direction and strength of association, calculated as the median SHAP difference between the upper (Q3) and lower (Q1) quartiles of each predictor.

**Global vs. Plant-Type-Specific Trait-Environment Relationships**

To further interpret how ecological parameters depend on climate, soil, and age predictors, we applied SHAP explainable AI analysis (Methods) to the spatialization neural network to isolate specific trait–environment relationships. When incorporating both plant functional type (PFT) classes and environmental variables, the combined PFT classes explain approximately 35–50% of the spatial variability, as measured by absolute SHAP values (Fig. S26), consistent with the hierarchical partition results (Fig. 2). To investigate the interaction between PFTs and environmental predictors, we compared functional relationships derived from model configurations with and without PFT predictors. The PFT-agnostic configuration (CLIM+SOIL+AGE) attempts to identify apparent global relationships between latent ecological parameters and spatial predictors, representing a hypothetical scenario where environment-trait relationships are uniform across plant types. Conversely, the PFT-aware model captures plant-type-specific dependencies between ecological parameters and environmental variables.

For a clearer extraction of environment-parameter relationships within each vegetation type, we restricted the SHAP analysis to pixels where a single plant type occupies at least 80% of the grid cell. Our analysis focused on four key ecological parameters that are robustly identifiable from global vegetation dynamics and exhibit distinct spatial patterns: photosynthetic efficiency, leaf carbon mass per area, carbon use efficiency, and root carbon allocation ratio. Among the environmental predictors, mean annual temperature, maximum canopy height, and—to a lesser extent—mean annual precipitation emerged as the most influential. We therefore examined their specific relationships with the selected ecological parameters (Fig. 4).

Mean annual temperature (MAP) exerts a dominant and often nonlinear influence on all four ecological parameters, indicating strong temperature regulation of ecosystem function consistent with the PC-level analysis. Photosynthetic efficiency rises with MAP from –15 °C to 15 °C before plateauing (Fig. 4a), while carbon use efficiency (NPP/GPP) declines from 0.45 to 0.35 and then stabilizes (Fig. 4g). Although global patterns are broadly mirrored within plant functional types (PFTs), notable differences exist in both mean values and temperature sensitivity. Grasslands and shrublands, spanning wide thermal ranges, generally exhibit higher photosynthetic efficiency (Fig. 3a), carbon use efficiency, (Fig. 4d) and root allocation ratios (Fig. 4j) than forests—traits reflecting acquisitive growth strategies in open habitats. In contrast, evergreen broadleaf forests show conservative strategies with low photosynthetic efficiency (Fig. 4a), high leaf carbon mass per area (Fig. 4d), and low carbon use efficiency Fig. 4g), adapted to humid tropical conditions. Deciduous broadleaf and evergreen needleleaf forests exhibit sharp internal temperature sensitivity despite lower mean efficiencies. These PFT-specific patterns highlight that plant-type-specific responses can diverge from global trends, underscoring the need to account for both within- and across-plant type variations in modeling trait–temperature relationships.

In contrast, the apparent global sensitivity of the four ecological parameters to mean annual precipitation largely reflects differences among distinct PFTs distributed along the precipitation gradient, rather than variations within PFTs. For example, grasslands and shrublands—typically characterized by higher photosynthetic efficiency—are more common in drier regions than forests, resulting in an apparent global decline in photosynthetic efficiency with increasing precipitation (Fig. 4b). However, within-PFT variation in precipitation is generally small. For carbon use efficiency (Fig. 4h) and root carbon allocation ratio, we still observe negative relationships with precipitation within most PFTs, consistent with global trends, though the within-PFT sensitivities are substantially smaller than the differences across PFTs. The divergence between global and within-PFT patterns is most pronounced in the relationship between maximum canopy height and photosynthetic capacity: while the global negative relationship is driven by differences in photosynthetic efficiency across PFTs (Fig. 4c), the within-PFT relationship is positive, suggesting that regions with greater resource availability support both higher canopy stature and photosynthetic potential. These contrasting patterns underscore the complementary roles of plant type and environmental gradients in shaping ecological parameter variation, reflecting distinct ecological processes operating at different organizational scales.

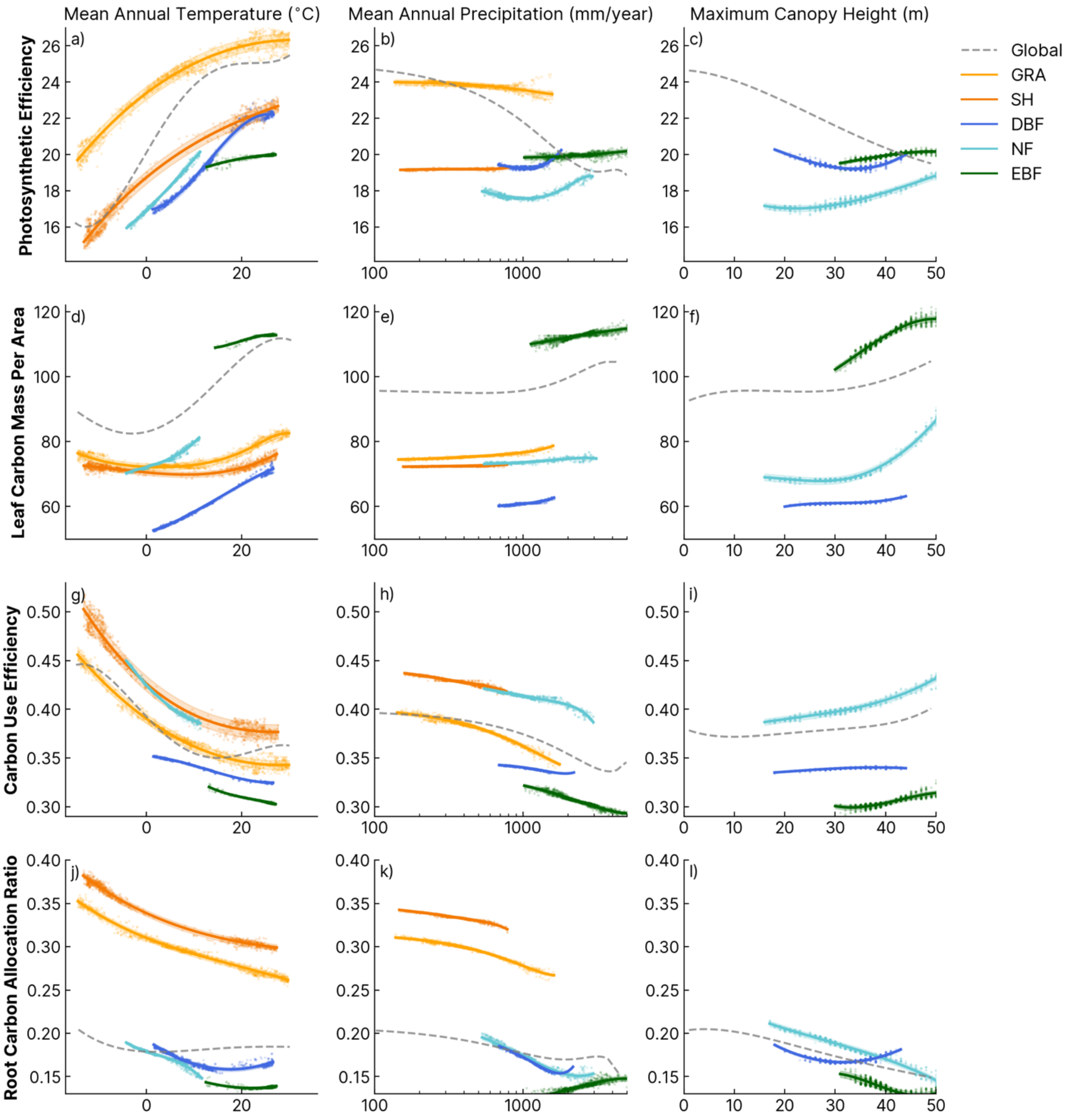

**Fig. 4 | Comparison of global and PFT-specific trait–environment relationships**. Panels show how nitrogen-limited photosynthetic efficiency (gC $m^{-2}$ leaf $day^{-1}$), leaf carbon mass per area (gC $m^{-2}$ leaf), carbon use efficiency (unitless, 0–1), and fine root carbon allocation ratio (unitless, 0–1) vary with mean annual temperature, mean annual precipitation, and maximum canopy height. The global relationships (dashed lines) represent ensemble-mean SHAP-derived dependencies from the CLIM+SOIL+AGE model configuration. PFT-specific relationships are derived from the PFT+CLIM+SOIL+AGE model configuration, using samples where the specified PFT comprises at least 80% of the pixel. PFT abbreviations: GRA – grassland; SH – shrubland; DBF – deciduous broadleaf forest; NF – needleleaf forest; EBF – evergreen broadleaf forest.

**Implications for Ecological Shifts Under Changing Climate**

The strong sensitivity of key ecological parameters to climatological gradients—particularly temperature—raises the question of whether spatial dependencies can inform potential shifts in ecological functioning under ongoing changes in mean climate. Assuming, as a theoretical extrapolation, that spatial-environmental relationships approximate temporal responses (see limitations of this assumption in the Discussions), whether ecological traits vary continuously along environmental gradients or reflect categorical differences among plant types has major implications for predicting terrestrial carbon and water dynamics in a changing climate. If ecological parameters within a given PFT show stronger sensitivity to temperature or precipitation than the global average, then ecosystems dominated by that PFT may exhibit amplified functional responses to similar climate shifts. Conversely, if ecological function is largely governed by PFT identity, with limited variation within types, then substantial functional changes would require shifts in vegetation composition—transitions that are often constrained by dispersal limits, recruitment bottlenecks, and potential ecological tipping points (e.g., a shift from tropical forest to savanna in the Amazon). While current PFT distributions reflect adaptation to past climate regimes over evolutionary timescales, they may act as ecological legacies that buffer against rapid climate-driven functional change on multi-decadal timescales relevant to contemporary climate change.

As an extrapolation experiment, we estimated theoretical changes in key ecological functional parameters using projected mean temperature and precipitation for 2081–2099 under the "Middle of the Road" SSP2-4.5 scenario[26], averaged across CMIP6 ensemble members (Table S5). These projections were evaluated using latent environment–parameter relationships derived from either a plant-type-agnostic model (CLIM+SOIL+AGE) or a plant-type-aware model (PFT+CLIM+SOIL+AGE) (Fig. S2). Under the plant-type-aware configuration, photosynthetic efficiency is projected to increase globally, with the largest gains occurring in boreal shrublands and temperate forests of the Northern Hemisphere due to both higher inferred sensitivity of photosynthetic efficiency to climatological temperature gradients in cooler regions and greater projected warming (Fig. S2a). However, incorporating plant-type-specific responses dampens the magnitude of this increase: the projected gain in photosynthetic efficiency is reduced by 0.5–1.75 gC $m^{-2}$ leaf $day^{-1}$ (about half of the magnitude) compared to estimates assuming a uniform global temperature dependency (Fig. S2b). Similarly, leaf carbon mass per area (LCMA) is projected to increase by 3-4 gC $m^{-2}$ in Northern Hemisphere temperate regions under the plant-type-agnostic model (Fig. S2b), but this increase is reduced by approximately 1-2 gC $m^{-2}$ when accounting for plant-type-specific effects (Fig. S2f). These results suggest that plant-type-specific constraints may reduce the magnitude of future increases in plant carbon uptake.

In contrast, both carbon use efficiency (Fig. S2c) and the root carbon allocation fraction (Fig. S2d) are projected to decline under the SSP2-4.5 scenario, leading to a lower NPP:GPP ratio and a potential shift in internal carbon allocation within plants across the Northern Hemisphere. While the projected decline in carbon use efficiency is similar under both the globally uniform and plant-type-aware relationships (Fig. S2g), plant-type-specific effects may amplify the reduction in root allocation (Fig. S2h)—resulting in a more pronounced decrease in the fraction of carbon allocated to fine roots, particularly in boreal shrublands and tundra ecosystems. These differences underscore the importance of accounting for functional diversity when projecting ecosystem responses to climate change. Nevertheless, we emphasize that these theoretical projections rely on a space-for-time substitution, assuming that current trait–environment relationships—reflecting equilibrium adaptations to climatological gradients—remain valid under future climate conditions. Whether such relationships will hold under anthropogenic climate change remains an open question, warranting further investigation through longitudinal studies and experimental evidence.

## Discussions

DifferLand demonstrates a scalable method for retrieving the dependencies of global ecological parameters to climate, age and soil. Unlike most model-data fusion studies that either optimize parameters independently at each grid cell [14,27] or rely on prescribed trait-environment functional forms for individual parameters [28], DifferLand captures the spatial coordination of ecological traits and their responses to environmental gradients through a unified latent space. This enables the derivation of a physically consistent, data-constrained set of ecological parameters that generalize better to unseen locations than traditional PFT-based approaches used in most current TBMs, thereby improving predictions of carbon and water fluxes across finer environmental gradients.

We note that the conclusions drawn from this research are dependent on the timeframe considered for the observations. In this 23-year study, spatial predictors were treated as temporally invariant, under the assumption that their temporal variations are negligible compared to their mean spatial gradients. Furthermore, we assumed that the relationships between parameters and the environment remained stable throughout the study period, implying that plant acclimation and adaptation to environmental changes occur on much longer timescales. However, these assumptions may not always hold for long-term predictions, in which case environmental conditions, land use patterns, population distributions[29], community compositions[30], and plant functional responses[31] could all undergo significant shifts. At multidecadal to centennial timescale, slow processes such $CO_2$ fertilization[32], nutrient limitation[33], evolving forest demography[34,35], and belowground carbon dynamics[36,37] can dominate the trajectories of land-atmosphere carbon exchange, yet most of these slow processes can be constrained only with long-term observations or by making equilibrium assumptions. In the absence of long-term global observational records on plant adaptation to climate change, the environment-parameter relationships learned from spatial gradients in this study offer insights into potential future shifts in ecosystem functions[38], assuming the validity of space-for-time substitution[39]. Future work should further investigate shifts in ecological parameters to changing environment using long-term observational records.

Analysis on identifiable model parameters unveiled correlations between them (Fig. 3). From a modeling perspective, inter-parameter correlation reduces the effective dimensionality of model parameters, and the sparsity of the latent parameter space is crucial for effective out-of-sample generalization in conditions unseen in the observational record[40–43]. Essentially, our ability to model high-dimensional physical worlds rely on the fact that most systems have much lower intrinsic dimensions governed by a few fundamental variables[40,44–47]. Natural selection, self-organization, and entropy maximization have been proposed as three organizing principles that introduce predictability to vegetation dynamics[48,49], giving rise to optimality-based trait spectra of photosynthesis[50,51], leaf size and economics[8,52–54], plant hydraulics[55–58], and carbon allocations[59–61] that reflect trade-offs under multiple selection forces.

Previous studies have leveraged dimensionality reduction algorithms on surface gas exchange measurements[7] and global trait databases[5,9] to extract the main axes of ecosystem traits and functions, yet they are limited by the available set of observable plant traits and uneven spatial sampling. By integrating the spatialization neural network with a differentiable TBM and assimilating diverse observations, DifferLand imposes a comprehensive set of constraints on the latent space, capturing both observed and unobserved dynamics and process dependencies. The three axes of latent parameter variations—leaf economics, plant stature, and agricultural intensity—highlight how these environment-parameter relationships influencing carbon and water fluxes emerge from the interactions of macroecological gradients and plant-type specific effects. It has been shown that state-of-the-art land models using PFT-based parameterizations overestimate the correlation between ecosystem functions compared with observations[7], limiting their capacity to simulate the full diversity of ecosystem function space, such as the spatial variations of carbon-use efficiency and water use strategies, and therefore likely their response to climate change[7]. Despite its parsimonious process representation (Text S7), DifferLand showcases how data-constrained hybrid differentiable modeling can be used to retrieve complex trait-environment relationships from observations to represent a more comprehensive view of the vegetation-environment relationships.

We propose that hybrid-physics modeling offers a promising pathway for parameter calibration in operational land surface models (LSMs), a process that is currently constrained by the high dimensionality of the parameter space and the substantial computational costs involved[62]. Many existing parameter estimation studies in LSMs have focused on only one or a few parameters at a time, often assigning uninformative and independent priors—represented by diagonal prior covariance matrices—which misrepresent potential parameter covariation and confounding effects[62]. By first identifying parameters that are both observationally constrained and climatically sensitive within an intermediate-complexity framework, researchers can prioritize a tractable subset of candidate parameters for subsequent analysis in full-complexity LSMs. Moreover, the robust spatial coordination of ecological traits revealed in our study provides a meaningful source of predictability to help address the "curse of dimensionality" in complex models[15,16]. By encoding these correlation relationships between parameters into an informative prior covariance matrix, we can robustly constrain the parameter search space for more complex models and reduce equifinality.

In the long term, the differentiable programming paradigm explored in DifferLand offers a promising avenue for addressing the structural and parametric biases that currently limit confidence in long-term projections of the terrestrial carbon sink[63]. Differentiable programming enables end-to-end learning of ecological functional relationships within an ecologically consistent modeling framework, creating new opportunities to uncover emergent patterns in ecosystem processes[64]. By integrating multi-modal observations of today's climate, ecological, and hydrological states, these emergent patterns can provide tighter constraints on uncertainties in future climate projections than traditional emergent-constraint studies, which largely depend on simple univariate linear relationships between current and projected climate[65]. Furthermore, coupling differentiable land models with differentiable general circulation models, such as NeuralGCM[66], opens a powerful future pathway for jointly learning biophysical and biogeochemical feedbacks between land and atmosphere across sub-seasonal to decadal timescales. While implementing a differentiable land model may require significant initial technical and numerical

investment—particularly in translating legacy components written in C or Fortran into modern differentiable frameworks—the potential long-term benefits are considerable. Such integration holds strong potential to improve the fidelity of future projections for both the terrestrial carbon cycle and the broader climate system in next-generation hybrid Earth system models[67].

## Methods

### DifferLand: A Hybrid-ML Terrestrial Biosphere Modeling Framework

The DifferLand configuration used in this study includes three main components: a spatialization neural network that maps spatial predictors to ecological model parameters, a process-based dynamical terrestrial biosphere model, and a loss function that computes the distance between simulated variables and observational constraints. We note that the DifferLand framework is flexible and it can accommodate different processes-based or neural network components to model ecological functional relationships[68]. The spatialization network ($f_{nn}$) (Fig. S3) consists of first a fully-connected neural network (FCNN) with 3 hidden-layers and 32 neurons that maps the input predictors ($\boldsymbol{P}$) to a 32-dimensional embedding. This embedding is then passed into three output layers, predicting the ecological model parameters of the Data Assimilation Linked Ecosystem Carbon (DALEC) model ($\boldsymbol{\theta}_e$, N=31), the initial pool values ($\boldsymbol{\theta}_i$ , N=7), and two phenology parameters ($\boldsymbol{\theta}_p$, N=2), respectively. $\boldsymbol{\theta}_e$ and $\boldsymbol{\theta}_p$ together constitute the model parameters of the DALEC model ($\boldsymbol{\theta}_M$). Rectified Linear Unit (ReLU) is used at the activation function of all NN layers. NN parameters ($\boldsymbol{\theta}_{nn}$) include weights and biases in each layer, with weights are randomly initialized using the Xavier initializer[69], with biases set to one. A transform function (Text S1) is used to convert each parameter from the real space to their physical range (see Table S2 for a list of the parameters and their physical range). In mathematical form (Equation 1), the spatialization network can be expressed as

$$\boldsymbol{\theta}_M^k, \boldsymbol{\theta}_i^k = f_{nn}\left(\boldsymbol{P}^{\boldsymbol{k}} \,\middle|\, \boldsymbol{\theta}_{nn}\right) \tag{1}$$

Where k denotes the kth pixel in spatial coverage. Note that $\boldsymbol{\theta}_{nn}$ is not spatially varying, as we assume a global relationship between environmental predictors and model parameters in this hierarchical framework.

The centerpiece of DifferLand is an automatically differentiable version of the Data Assimilation Linked Ecosystem Carbon (DALEC) model, meaning that the model has the capacity to compute gradients and the sensitivity to any parameter or variable in the model with respect to a model output using back-propagation (adapted from model version 1006, see Fig. S5 for model schematics). DALEC is an intermediate-complexity dynamical terrestrial biosphere model that simulates photosynthesis, carbon allocation, leaf phenology, autotrophic and heterotrophic respiration, turnover, decomposition, and fire disturbance. Labile, foliar, wood, fine roots, litter, and soil carbon pools are prognostically computed at each time step based on mass balance principles[14,70,71]. The model also simulates ET based on an underlying water use efficiency formulation[72,73], and computes a prognostic water bucket as a balance between precipitation, runoff, and ET. The water bucket feedbacks into photosynthesis to represent soil water limitation on GPP[68,74]. Versions of the DALEC model have been used in the CARbon DAta MOdel fraMework (CARDAMOM) to conduct MDF studies at local[70,75,76], regional[74,77], to global scales[14]. At each time step t for location k, this TBM ($f_M$) uses meteorological forcing variables ($\boldsymbol{m}^{k,t}$) to prognostically evolve state variables ($\boldsymbol{x}$) and compute observable ($\overline{\boldsymbol{y_o}}$) and unobservable ($\overline{\boldsymbol{y_u}}$) ecological variables and fluxes (Equation 2). We note that at t=0, $\boldsymbol{x}^{k,0} = \boldsymbol{\theta}_i$

$$\widehat{\boldsymbol{y}_o^{k,t}}, \widehat{\boldsymbol{y}_u^{k,t}}, \boldsymbol{x}^{k,t+1} = f_M\left(\boldsymbol{x}^{k,t}, \boldsymbol{m}^{k,t} \,\middle|\, \boldsymbol{\theta}_M^k\right) \quad (2)$$

Substituting (1) into (2) and considering the legacy effects of meteorological forcing on ecological states, we have

$$\widehat{\boldsymbol{y}_o^{k,t}}, \widehat{\boldsymbol{y}_u^{k,t}}, \boldsymbol{x}^{k,t+1} = f_M\left( \boldsymbol{m}^{k,1}, \boldsymbol{m}^{k,2}, \ldots, \boldsymbol{m}^{k,t} \middle| f_{nn}\left(\boldsymbol{P}^{\boldsymbol{k}} \,\middle|\, \boldsymbol{\theta}_{nn}\right) \right) = f_M\left( \boldsymbol{m}^{k,1}, \boldsymbol{m}^{k,2}, \ldots, \boldsymbol{m}^{k,t} \middle| \boldsymbol{P}^{\boldsymbol{k}}, \boldsymbol{\theta}_{nn}\right) \quad (3)$$

Equation 3 essentially states that the ecological fluxes at states any time and location is a function of the environmental background ($\boldsymbol{P}^{\boldsymbol{k}}$), the environment-parameter relationships learned through the spatialization network ($\boldsymbol{\theta}_{nn}$), the current and legacy meteorological forcings up to that time ($\boldsymbol{m}^{k,1}, \boldsymbol{m}^{k,2}, \ldots, \boldsymbol{m}^{k,t}$), and the ecological process dependencies encoded in the TBM ($f_M$).

By selecting a parsimonious TBM, we aim to include mechanistic representations that capture essential ecological processes while minimizing equifinality and the high computational costs associated with more sophisticated models. Despite its relative simplicity, DALEC has shown strong performance in capturing global carbon flux dynamics within ranges comparable to much more complex Dynamic Global Vegetation Models (DGVMs) and fully-coupled Earth System Models (ESMs)[78]. A brief summary of the DALEC model can be found in Text S2, and additional details about the adaptations made to convert DALEC into a differentiable model has been described in previous work[68]. We present a worked example from a site in Finland illustrating the inner mechanism of the DifferLand framework: the spatialization network maps environmental predictors into ecological parameters, which then parameterize the DALEC model to simulate trajectories of ecological states and fluxes under monthly meteorological forcing (Fig. S4).

The loss function (equation 4) calculates a weighted sum of mean squared error (MSE) between model simulated ($\widehat{\boldsymbol{y}_o}$) and observed ($\boldsymbol{y}_o$) NEE, LAI, SIF, ET, RECO, VOD, live biomass, fire C emission, and soil organic carbon from gridded datasets.  It also compares the differences between simulated and observed GPP, RECO, and ET at a global network of eddy covariance sites where grid cells overlap with tower sites (Fig. S4). Among the variables that are not directly simulated by DALEC but assimilated to inform carbon dynamics, (reconstructed) SIF is assumed to have a linear relationship between modeled GPP at monthly level[79], with the slope and intercept of the GPP-SIF relationships treated as model parameters to be predicted from the spatialization neural network[74].

Live biomass was computed as the sum of the labile, foliar, wood, and fine roots carbon pools. In addition, we included two soft constraint terms ($\mathcal{L}_{constraint}$) to prevent excessive drying of the water pool and penalize spurious exponential growth and decay in carbon and water pools, as this was previously found important for predicting water dynamics in semi-arid regions and prevent excessive excursions in carbon pool trajectories due to implausible initial conditions (Text S3)[68].

Thus, we have

$$\mathcal{L} = \sum_{k,t,v} \alpha_v \left( \widehat{\boldsymbol{y}_{\boldsymbol{o}}^{k,t,v}} - \boldsymbol{y}_{\boldsymbol{o}}^{k,t,v} \right)^2 + \mathcal{L}_{constraint} \tag{4}$$

Index v denotes the v[th] variable in the loss function. The gradient of the loss function with respect to the NN parameters in the spatialization network ($\boldsymbol{\theta}_{nn}$) is used to optimize the entire framework. We heuristically tuned the weights for different loss terms ($\alpha_v$) to achieve a balanced performance on all observational constraints. The complete set of hyperparameters used in this study are listed in Table S3. We implemented DifferLand in JAX[80], an automatic differentiation software package built in Python. All code and data used in this study can be accessed with the links provided in the Code & Data Availability sections.

**Datasets**

DifferLand requires three types of data inputs: spatial predictors ($\boldsymbol{P}$), forcing variables ($\boldsymbol{m}$), and observational constraints ($\boldsymbol{y_o}$).

*Spatial predictors*

Spatial variables are classified into four groups: PFT, CLIM, AGE, and SOIL (i.e., $\boldsymbol{P} \subseteq \{PFT, CLIM, SOIL, AGE\}$). PFT variables are derived from MCD12C1.v061 MODIS/Terra + Aqua Yearly Land Cover Type[81], with the percentage of underlying 500m classes aggregated to 0.25° grid cells. We consolidated the International Geosphere Biosphere Programme (IGBP) fractions into 11 classes including needleleaf forest (ENF+DNF), deciduous broadleaf forest (DBF), evergreen broadleaf forest (EBF), mixed forest (MF), shrubland (SH), savanna (WSA+OSA), grassland (GRA), wetland (WET), cropland (CRO+CNM), and a non-vegetated land surface class encompassing all other land cover types (URB+SNO+BSV). Pixels where the area of permanent water bodies exceeds 50% or non-vegetated surface exceeds 20%, were excluded from the analysis.

The CLIM group includes mean annual temperature (MAT, °C) and mean annual precipitation (MAP, mm/year) averaged over 2001-2023 from ERA5 reanalysis by ECMWF[82]. We also elevation (ELE, m) derived from the Global 30 Arc-Second Elevation dataset (GTOPO30)[83,84].

The AGE predictor group includes forest age (year) estimates obtained from a global 1 km forest age dataset (circa 2010)[85,86], which used a machine learning algorithm and data from more than 40,000 forest inventory plots to estimate global forest age. We used tree age estimates with 10% tree cover correction for our analysis, whereas non-forest pixels were assigned an age value of 1 year assuming annual vegetation turnover. Also in this group is the maximum canopy height (CAN.H, m) from the ETH Global Sentinel-2 10 m Canopy Height dataset[87], accessed from Google Earth Engine. We first computed the $95^{th}$-percentile canopy height at 1 km spatial resolution to represent forest growth potential while avoiding single pixel outliers, followed by calculating maximum value among the 1 km pixels within each 0.25° grid cell.

SOIL predictors include soil bulk density and soil sand, silt, clay, and gravel fractions derived from the Regridded Harmonized World Soil Database v1.2[88] available at 0.05° resolution. We combined surface soil (0-30cm) and deeper soil values (30-100cm) to a top 1m value (0-100cm) by computing a depth weighted mean. All spatial predictors are assumed to be temporally invariant and converted to a common spatial resolution of 0.25°.

*Forcing data*

Monthly-averaged forcing data used to drive the DALEC model include daily min temperature (°C), daily max temperature (°C), shortwave solar radiation downward ($W/m^2$), precipitation rate (mm/day), and VPD (kPa) computed from ERA5 Reanalysis[82] between 2001-2023 at 0.25° resolution. Globally-averaged monthly $CO_2$ concentration (ppm) was obtained from NOAA Global Monitoring Laboratory measurements[89]. Fire dynamics were driven by burned area from the Fifth Version of Global Fire Emissions Database (GFED5)[90] at 0.25° resolution between 2001 and 2020. We divided the burned area by the burnable area estimated in GFED5 to obtain burned area fraction. Because GFED5 burned area data are not yet fully available after 2020, we cross-calibrated GFED5 with monthly MODIS burned area fraction data (MCD43A1.v061) resampled to 0.25° to extend the burned area fraction record through 2021–2023.

*Observational constraints*

We assimilated 12 globally gridded or in-situ datasets to constrain various aspects of the terrestrial biosphere represented in DifferLand (Fig. S2). We used 0.05° biweekly MODIS-based Long-term Continuous SIF-informed Photosynthesis Proxy (LCSPP-MODIS) as a proxy of photosynthesis between 2010-2023[91,92]. Reprocessed MODIS Version 6.1 Leaf Area Index[93,94], which exhibited enhanced spatial and temporal continuity compared with the original MCD15A2H, was used to constrain modeled foliar dynamics from 2001-2023. Evapotranspiration from Global Land Evaporation Amsterdam Model (GLEAM) v4.2a at 0.1° resolution was used to constrain modeled ET between 2001-2023[24]. Monthly GFED5 (beta) total fire carbon emission at 0.25° is used to constrain modeled carbon emission. The aforementioned global datasets were regirded to 0.25° grid cells at monthly intervals to match DifferLand output.

Annual live woody biomass ($B_w$) maps from 2001-2019 were obtained from a dataset reconstructed with an array of inventory plots, airborne, and satellite observations[95]. Herbaceous biomass ($B_h$) is estimated under an equilibrium assumption with $B_h = f_h \times \overline{GPP}/\tau \times (1 - \alpha)$, where $f_h$ is the area fraction of herbaceous vegetation within each grid cell estimated from MCD12C1, $\overline{GPP}$ is annual GPP climatology estimated from FLUXCOM upscaling[96], $\alpha$ is respiration cost of carbon (assumed to be 0.5), and $\tau$ is the mean residence time taken as one year for annual plants[97]. Total live biomass ($B_l$) is computed as $B_l=B_h+B_w$. In the absence of temporal observations, this use of equilibrium assumption provides an estimate of the spatial gradient of live biomass density. The lack of temporal variability in biomass estimates is partially compensated by assimilated LAI and VOD dynamics in those regions. We aggregated annual live biomass density to a 0.25° spatial resolution.

Monthly NBE from the NASA Carbon Monitoring System Flux (CMS-Flux) GCP 2023 submission[22,25,98] was used to constrain DifferLand modeled NBE from 2010-2022. CMS-Flux is a top-down flux inversion system constrained by column $CO_2$ observations from the Greenhouse Gases Observing Satellite (GOSAT) and Observing Carbon Observatory-2 (OCO-2), which have broader and more even spatial coverage compared with ground-based $CO_2$ observations[99,100]. Due to the computational cost of the atmospheric transport model, CMS-Flux has a native resolution of 4°×5°. We derived NBE in DifferLand as the sum of (negative) gross primary productivity (GPP), ecosystem respiration (RECO, which is itself the sum of autotrophic respiration, $R_a$, and heterotrophic respiration, $R_h$) and carbon emission from fire ($F_{fire}$). The first three terms constitute net ecosystem exchange (NEE), whereas fire carbon emission was additionally constrained by the GFED5 dataset. A negative NBE represents a net flux of carbon from the atmosphere to the land (i.e., land carbon sink).

$$NBE = -GPP + R_a + R_h + F_{fire} = -GPP + RECO + F_{fire} = NEE + F_{fire}$$

To constrain DifferLand simulated terrestrial water storage, we assimilated monthly satellite-gravimetry based JPL GRACE and GRACE-FO Mascon Equivalent Water Height (GRACE EWT) from 2002-2023 (RL06.3)[23]. Although GRACE EWT was provided on 0.5° grids, the product has an effective resolution of 3°×3°. We thus assimilated both CMS-Flux NBE and GRACE EWT at the 4°×5° batch level to prevent signal aliasing (see the next section for details). To evaluate the robustness of model performance to different observational constraints, we conducted sensitivity analyses with alternative assimilated datasets, including LAI, live biomass, NBE, fire carbon emissions, and optimally assimilated VOD. Detailed results are presented in Text S4.

We also assembled monthly eddy-covariance based observations GPP, RECO, and ET by combining the FLUXNET2015[101], ICOS, OzFlux[102], and AmeriFlux FLUXNET datasets. For GPP and RECO, we used night-time partitioned values with variable $u^*$ threshold and removed months where more than 30% of the NEE measurements were gap-filled[103]. For ET, we removed months where more than 30% of the latent heat flux measurements were gap-filled. Sites representing managed landscapes that are highly uncharacteristic of the surrounding grid cell ecosystems are excluded from the data assimilation. After filtering, we retained 180 sites with more than 12 months of observations for all three variables during the simulation period (Fig. S33). Whenever available, we gridded the site-level data to 0.25°. If observations from multiple sites within a grid cell were available in a specific month, we used the mean value across these sites to fill the grid cell. To investigate the contributions of eddy covariance data and potential uncertainties due to spatial mismatch between model grid and tower footprints, we further performed a set of sensitivity analyses by either not assimilating eddy covariance data or testing the effects of more stringent or relaxed site representativeness filtering criteria (Text S5). We found that while assimilating eddy covariance data is essential for constraining both the absolute magnitude and latitudinal gradient of GPP and RECO, the results are largely insensitive to the specific thresholds used for site representativeness filtering (Fig. S35). A complete list of the eddy covariance sites used in this study is provided in SI (Table S4).

**Model Training & Evaluation**

To assimilate multi-resolution observational constraints at both fine (0.25°) and coarse (4°x5°) resolutions, we first divided the globe into 3240 4°×5° patches, each corresponding to the grid of coarse-resolution datasets. Each 4°×5° patch contains 320 nested fine-resolution grid cells (Fig. S6). We filtered these patches to retain only those with at least 32 valid vegetated fine-resolution grid cells. Out of the 944 patches meeting this criterion, 10% of the patches (N = 95) were randomly selected and reserved for model testing (Fig. S6c). Prior to training each ensemble member, we randomly sampled 90% of the remaining 849 patches for training, while the unselected patches formed a development set used for hyperparameter tuning and monitoring training progress (Fig. S6a). The final training and development sets contain 164,152 grid cells with a total of 45,305,852 pixel-months. All spatial predictors were standardized by subtracting their means and dividing by their standard deviations, calculated from the training dataset, to accelerate model convergence. After evaluating model performance, we retrained each ensemble on the combined training and development sets (excluding the test set) to derive latent ecological parameters and their relationships with spatial predictors for subsequent analysis.

We developed a customized stochastic gradient descent algorithm to train the multi-resolution model. This approach first permutes the fine-resolution grids within each patch and then shuffles the patches within the training set to introduce stochasticity, while ensuring that all fine-resolution grid cells within a patch remain grouped. As a result, each patch forms a minibatch (batch size = 320) during model training, allowing fine-resolution datasets to serve as sample-level constraints, and coarse-resolution datasets as batch-level constraints. We assimilate coarse resolution constraints at the patch level only if at least 80% of the fine resolution grid cells within the patch are valid. The model was trained using the Adam optimizer, with a learning rate of 0.0005. Each ensemble member underwent 199 epochs of training, a choice balanced between computational cost, model convergence, and mitigating overfitting risks.

After training, we evaluated model performance on the held-out test datasets. We used $R^2$-score as a measure of overall model fitting on each variable

$$R^2 = 1 - \frac{\sum_{i=1}^{n}(y_i - \hat{y_i})^2}{\sum_{i=1}^{n}(y_i - \bar{y})^2}$$

With $y_i$ being $i^{th}$ observed value, $\hat{y_i}$ being the $i^{th}$ modeled value, and $\bar{y}$ is the observed mean. The $R^2$-score is mathematically equivalent to the Nash-Sutcliffe Efficiency (NSE) metric commonly used in hydrology, which accounts for both correlation and systematic offsets between modeled and observed values. A perfect model would have $R^2$=1, but the $R^2$-score can approach -Inf for arbitrarily bad predictions. Thus, we also used the explained variance score ($r^2$), computed as the square of the Pearson correlation coefficient between modeled and observed values (range 0-1), for hierarchical partition and pixelwise temporal correlations. We made explicit in our manuscript which metric was used when reporting the results. We excluded the first two years of simulations (2001–2002) to reduce the potential influence of initialization uncertainties on performance metrics.

**Hierarchical Partition of Explained Variance**

Hierarchical partition[104] enables a decomposition of the explained variance on target variables by a multivariate regression model into the independent contributions of different groups of variables. To segregate the independent contributions of PFT, CLIM, SOIL, and AGE, we performed a full factorial experiment by including or excluding each group of variables, resulting in a total of 15 setups (i.e., $|P(\{PFT, CLIM, SOIL, AGE\})-\{\emptyset\}|=2^4-1$, where P (·) denotes power set).

For each setup, we ran 20 experiments with random initializations and selected the 10 with the lowest training loss, minimizing the impact of poor initializations and forming a robust model ensemble. The mean performance metric across these 10 ensemble members is used to represent a setup. Under the hierarchical theorem, the independent effect of a variable group j within a set containing N groups of variables can be computed as the mean explained variance difference between N pairs of setups within a predictor group hierarchy where j is either included or excluded from the predictor set, averaged over all possible (N-1)! sequences of hierarchies[104]. We conducted a hierarchical partitioning analysis on explained variance for each of SIF, LAI, ET, NEE, and EWT on both global and pixel levels. We then derived the proportion of explained variance that can be independently attributed to each predictor group, assuming a baseline explained variance of 0.

**Parameter Identifiability Analysis**

We conducted parameter identifiability analyses on the three setups using either the full set of predictors (PFT+CLIM+SOIL+AGE, Fig. S36), the environmental predictors only (CLIM+SOIL+AGE, Fig. S37), or the PFT predictors only (PFT, Fig. S38). For each model parameter and initial pool value, we computed the pixelwise coefficient of variance (CV) across ensemble members and used the spatial median CV as a diagnostic for predictor robustness. The rationale of this test is that if a latent ecological parameter is well constrained by the assimilated observations, it should converge to relatively stable values across independently initialized runs. Conversely, non-identifiable parameters with low sensitivity to the observational constraints can be expected to take substantially different values across independently initialized runs. We used a spatial median CV $< 0.6$ to select parameters for subsequent analyses to balance parameter robustness with process coverage under the available constraints.

**Principal Component Analysis (PCA) on Ecological Parameters**

We performed a principal component analysis (PCA) on 13 ecologically significant and spatially consistent parameters to determine the underlying dimensions controlling their spatial variability. These parameters include canopy photosynthetic efficiency (ce), underlying water use efficiency (uWUE), temperature sensitivity of heterotrophic respiration (Q10), autotrophic respiration fraction ($A_{auto}$), leaf carbon mass per area (LCMA), leaf onset ($c_{ronset}$) and fall duration ($c_{rfall}$), leaf lifespan ($\tau_{foliar}$), and allocation fraction of GPP to labile ($A_{labile}$), foliar ($A_{foliar}$), wood ($A_{wood}$), and fine roots ($A_{root}$) pools. Overall, these parameters characterize ecosystem functions related to photosynthesis & leaf phenology, carbon allocation, residence time, and water use efficiency. We projected these parameters into the space defined by the first three principal components to explore their interrelationships (Fig. 3)

### SHAP Analysis for Feature Importance & Environment-Parameter Relationships

We applied kernel SHapley Additive exPlanations (SHAP) to extract the learned environment-parameter relationship from the spatialization neural network for a selection of latent ecological parameters[105]. Kernel SHAP estimates Shapley values, informed by cooperative game theory, to quantify each feature's marginal and additive contribution to the model's output. To ensure robustness of SHAP results and reduce the random uncertainties associated with individual model members, we calculated ensemble-based SHAP by first defining an ensemble-averaged model $M_{ens}$ from 10 out of 20 model members for each configuration that best converges on the training dataset.

$$M_{ens} = \frac{1}{10}\sum_{i=1}^{10} M(x, \theta_i)$$

Where M is the spatialization network, x denotes the spatial predictions, and $\theta_i$ are the neural network weights and biases of member i within the ensemble. From $M_{ens}$ we sampled 100 grid cells to compute the background distribution, and then computed the SHAP values from 1,000 randomly selected grid cells within the training dataset. Feature importance of different predictors was determined by ranking the mean absolute SHAP values from the 1,000 samples. This procedure is repeated for each selected parameter to obtain global trait-environment relationships. To derive plant-type-specific SHAP results, we further computed conditional SHAP values by sampling from grid cells where a specified plant functional type constitutes at least 80% of the area. For SHAP analysis on the principal components, we linearly projected $M_{ens}$ into the principal component space using the PCA loadings and computed the SHAP analysis within this projected space.

## Data Availability

The driver files used for the DifferLand model are stored at doi:10.5281/zenodo.13984225.

## Code Availability

The code of the DifferLand model will be made public at publication.

## Conflicts of Interests

The authors declare no competing interests.

## Acknowledgements

We would like to thank Weiwei Zhan for providing valuable comments that helped us improve the manuscript. We would also like to express our appreciation for Dr. Nuno Carvalhais and members of the Model-Data Integration group at the Max Planck Institute for Biogeochemistry (MPI-BGC) for the insightful discussions related to this work during J.F.'s academic visit. J.F. and P.G. are supported by the NASA ROSES-22 Future Investigators in NASA Earth and Space Science and Technology (FINESST) Program (Grant Number: 80NSSC24K0023). This work was also funded by the Land Ecosystem Models based On New Theory, obseRvations, and ExperimEnts (LEMONTREE) project (UREAD 1005109-LEMONTREE) through the generosity of Eric and Wendy Schmidt by recommendation of the Schmidt Futures programme, the European Research Council grant USMILE (ERC CU18-3746), the Max-Planck Caltech Carnegie Columbia (MC$^3$) Center funded by the Max Planck Foundation, and National Science Foundation Science and Technology Center LEAP, Learning the Earth with Artificial intelligence and Physics (AGS-2019625). Funding for the AmeriFlux data portal was provided by the U.S. Department of Energy Office of Science. Fig. 1 was created in BioRender. Fang, J. (2026) https://BioRender.com/ne0dr93

## References

1. Cadotte, M. W. & Tucker, C. M. Should Environmental Filtering be Abandoned? *Trends in Ecology & Evolution* **32**, 429–437 (2017).
2. Kraft, N. J. B. *et al.* Community assembly, coexistence and the environmental filtering metaphor. *Functional Ecology* **29**, 592–599 (2015).
3. Meinzer, F. Functional convergence in plant responses to the environment. *Oecologia* **134**, 1–11 (2003).
4. Anderegg, L. D. L. Why can't we predict traits from the environment? *New Phytologist* **237**, 1998–2004 (2023).
5. Díaz, S. *et al.* The global spectrum of plant form and function. *Nature* **529**, 167–171 (2016).
6. Bruelheide, H. *et al.* Global trait–environment relationships of plant communities. *Nat Ecol Evol* **2**, 1906–1917 (2018).
7. Migliavacca, M. *et al.* The three major axes of terrestrial ecosystem function. *Nature* (2021) doi:10.1038/s41586-021-03939-9.
8. Wright, I. J. *et al.* Global climatic drivers of leaf size. *Science* **357**, 917–921 (2017).
9. Joswig, J. S. *et al.* Climatic and soil factors explain the two-dimensional spectrum of global plant trait variation. *Nat Ecol Evol* **6**, 36–50 (2021).
10. Famiglietti, C. A. *et al.* Global net biome CO2 exchange predicted comparably well using parameter–environment relationships and plant functional types. *Global Change Biology* **29**, 2256–2273 (2023).
11. Cranko Page, J., Abramowitz, G., De Kauwe, Martin. G. & Pitman, A. J. Are Plant Functional Types Fit for Purpose? *Geophysical Research Letters* **51**, e2023GL104962 (2024).
12. Butler, E. E. *et al.* Mapping local and global variability in plant trait distributions. *Proc. Natl. Acad. Sci. U.S.A.* **114**, (2017).
13. Liu, Y., Holtzman, N. M. & Konings, A. G. Global ecosystem-scale plant hydraulic traits retrieved using model–data fusion. *Hydrology and Earth System Sciences* **25**, 2399–2417 (2021).
14. Bloom, A. A., Exbrayat, J.-F., Van Der Velde, I. R., Feng, L. & Williams, M. The decadal state of the terrestrial carbon cycle: Global retrievals of terrestrial carbon allocation, pools, and residence times. *Proceedings of the National Academy of Sciences* **113**, 1285–1290 (2016).
15. Xu, X. & Trugman, A. T. Trait-Based Modeling of Terrestrial Ecosystems: Advances and Challenges Under Global Change. *Curr Clim Change Rep* **7**, 1–13 (2021).
16. Fisher, R. A. & Koven, C. D. Perspectives on the Future of Land Surface Models and the Challenges of Representing Complex Terrestrial Systems. *Journal of Advances in Modeling Earth Systems* **12**, e2018MS001453 (2020).
17. Fan, H. *et al.* Physically Consistent Global Atmospheric Data Assimilation with Machine Learning in Latent Space. Preprint at https://doi.org/10.48550/arXiv.2502.02884 (2025).
18. Brown, C. F. *et al.* AlphaEarth Foundations: An embedding field model for accurate and efficient global mapping from sparse label data. Preprint at https://doi.org/10.48550/arXiv.2507.22291 (2025).
19. Bengio, Y., Courville, A. & Vincent, P. Representation Learning: A Review and New Perspectives. *IEEE Trans. Pattern Anal. Mach. Intell.* **35**, 1798–1828 (2013).
20. Knyazikhin, Y. *et al.* MODIS Leaf Area Index (LAI) and Fraction of Photosynthetically Active Radiation Absorbed by Vegetation (FPAR) Product (MOD15) Algorithm Theoretical Basis Document. (1999).
21. Fang, J. *et al.* A long-term reconstruction of a global photosynthesis proxy over 1982–2023. *Sci Data* **12**, 372 (2025).
22. Liu, J. CMS-Flux GCP 2023 Submission. (2024).
23. Wiese, D. N., Yuan, D.-N., Boening, C., Landerer, F. W. & Watkins, M. M. JPL GRACE and GRACE-FO Mascon Ocean, Ice, and Hydrology Equivalent Water Height CRI Filtered. Physical Oceanography Distributed Active Archive Center https://doi.org/10.5067/TEMSC-3JC63 (2023).

24. Miralles, D. G. *et al.* GLEAM4: global evaporation and soil moisture datasets at 0.1° resolution from 1980 to near present.
25. Friedlingstein, P. *et al.* Global carbon budget 2023. *Earth System Science Data* **15**, 5301–5369 (2023).
26. O'Neill, B. C. *et al.* The Scenario Model Intercomparison Project (ScenarioMIP) for CMIP6. *Geosci. Model Dev.* **9**, 3461–3482 (2016).
27. Smallman, T. L. *et al.* Parameter uncertainty dominates C-cycle forecast errors over most of Brazil for the 21st century. *Earth System Dynamics* **12**, 1191–1237 (2021).
28. Robinett, T. W. *et al.* Parameterizing stomatal conductance based on trait-environment relationships often improves land surface model predictions of evapotranspiration and streamflow. Preprint at https://doi.org/10.22541/essoar.174139370.00258712/v1 (2025).
29. Kelly, A. E. & Goulden, M. L. Rapid shifts in plant distribution with recent climate change. *Proc. Natl. Acad. Sci. U.S.A.* **105**, 11823–11826 (2008).
30. Liu, H. *et al.* Shifting plant species composition in response to climate change stabilizes grassland primary production. *Proc. Natl. Acad. Sci. U.S.A.* **115**, 4051–4056 (2018).
31. Franks, S. J., Weber, J. J. & Aitken, S. N. Evolutionary and plastic responses to climate change in terrestrial plant populations. *Evolutionary Applications* **7**, 123–139 (2014).
32. Quetin, G. R. *et al.* Attributing past carbon fluxes to $CO_2$ and climate change: Respiration response to $CO_2$ fertilization shifts regional distribution of the carbon sink. *Global Biogeochemical Cycles* (2023) doi:10.1029/2022GB007478.
33. Terrer, C. *et al.* Nitrogen and phosphorus constrain the CO2 fertilization of global plant biomass. *Nat. Clim. Chang.* **9**, 684–689 (2019).
34. Robinson, N. *et al.* Protect young secondary forests for optimum carbon removal. (2024).
35. Yang, H. *et al.* Global increase in biomass carbon stock dominated by growth of northern young forests over past decade. *Nat. Geosci.* **16**, 886–892 (2023).
36. Huang, Y. *et al.* Size, distribution, and vulnerability of the global soil inorganic carbon. *Science* **384**, 233–239 (2024).
37. Tao, F. *et al.* Microbial carbon use efficiency promotes global soil carbon storage. *Nature* (2023) doi:10.1038/s41586-023-06042-3.
38. Space-for-Time Substitution as an Alternative to Long-Term Studies. in *Long-Term Studies in Ecology* 110–135 (Springer New York, New York, NY, 1989). doi:10.1007/978-1-4615-7358-6_5.
39. Damgaard, C. A Critique of the Space-for-Time Substitution Practice in Community Ecology. *Trends in Ecology & Evolution* **34**, 416–421 (2019).
40. Chen, B. *et al.* Automated discovery of fundamental variables hidden in experimental data. *Nat Comput Sci* **2**, 433–442 (2022).
41. Floryan, D. & Graham, M. D. Data-driven discovery of intrinsic dynamics. *Nat Mach Intell* **4**, 1113–1120 (2022).
42. Brunton, S. L., Proctor, J. L. & Kutz, J. N. Discovering governing equations from data by sparse identification of nonlinear dynamical systems. *Proc. Natl. Acad. Sci. U.S.A.* **113**, 3932–3937 (2016).
43. Champion, K., Lusch, B., Nathan Kutz, J. & Brunton, S. L. Data-driven discovery of coordinates and governing equations. *Proceedings of the National Academy of Sciences of the United States of America* **116**, 22445–22451 (2019).
44. Famiglietti, C. A. *et al.* Optimal model complexity for terrestrial carbon cycle prediction. *Biogeosciences* **18**, 2727–2754 (2021).
45. Buckingham, E. On physically similar systems; illustrations of the use of dimensional equations. *Physical review* **4**, 345 (1914).
46. Porporato, A. Hydrology without dimensions. *Hydrol. Earth Syst. Sci.* **26**, 355–374 (2022).
47. Feng, X. *et al.* The ecohydrological context of drought and classification of plant responses. *Ecology Letters* **21**, 1723–1736 (2018).
48. Franklin, O. *et al.* Organizing principles for vegetation dynamics. *Nat. Plants* **6**, 444–453 (2020).

49. Harrison, S. P. *et al.* Eco-evolutionary optimality as a means to improve vegetation and land-surface models. *New Phytologist* **231**, 2125–2141 (2021).
50. Prentice, I. C., Dong, N., Gleason, S. M., Maire, V. & Wright, I. J. Balancing the costs of carbon gain and water transport: testing a new theoretical framework for plant functional ecology. *Ecology Letters* **17**, 82–91 (2014).
51. Joshi, J. *et al.* Towards a unified theory of plant photosynthesis and hydraulics. *Nat. Plants* **8**, 1304–1316 (2022).
52. De La Riva, E. G. *et al.* A plant economics spectrum in Mediterranean forests along environmental gradients: is there coordination among leaf, stem and root traits? *J Vegetation Science* **27**, 187–199 (2016).
53. Dong, N., Dechant, B., Wang, W., Wright, I. J. & Prentice, I. C. Global leaf-trait mapping based on optimality theory. *Global Ecol Biogeogr* **32**, 1152–1162 (2023).
54. Wang, H. *et al.* Leaf economics fundamentals explained by optimality principles. *Sci. Adv.* **9**, eadd5667 (2023).
55. Xu, H., Wang, H., Prentice, I. C., Harrison, S. P. & Wright, I. J. Plant hydraulics coordinated with photosynthetic traits and climate. Preprint at https://doi.org/10.1101/2021.03.02.433324 (2021).
56. Anderegg, W. R. L. *et al.* Woody plants optimise stomatal behaviour relative to hydraulic risk. *Ecology Letters* **21**, 968–977 (2018).
57. Mencuccini, M., Minunno, F., Salmon, Y., Martínez-Vilalta, J. & Hölttä, T. Coordination of physiological traits involved in drought-induced mortality of woody plants. *New Phytologist* **208**, 396–409 (2015).
58. Franklin, O., Fransson, P., Hofhansl, F., Jansen, S. & Joshi, J. Optimal balancing of xylem efficiency and safety explains plant vulnerability to drought. *Ecology Letters* **26**, 1485–1496 (2023).
59. De La Riva, E. G. *et al.* The Economics Spectrum Drives Root Trait Strategies in Mediterranean Vegetation. *Front. Plant Sci.* **12**, 773118 (2021).
60. Schymanski, S. J., Sivapalan, M., Roderick, M. L., Beringer, J. & Hutley, L. B. An optimality-based model of the coupled soil moisture and root dynamics. *Hydrology and Earth System Sciences* **12**, 913–932 (2008).
61. Gentine, P., D'Odorico, P., Lintner, B. R., Sivandran, G. & Salvucci, G. Interdependence of climate, soil, and vegetation as constrained by the Budyko curve. *Geophysical Research Letters* **39**, (2012).
62. Raoult, N. *et al.* Parameter Estimation in Land Surface Models: Challenges and Opportunities with Data Assimilation and Machine Learning. Preprint at https://doi.org/10.22541/essoar.172838640.01153603/v1 (2024).
63. Bonan, G. B. & Doney, S. C. Climate, ecosystems, and planetary futures: The challenge to predict life in Earth system models. *Science* **359**, eaam8328 (2018).
64. Shen, C. *et al.* Differentiable modelling to unify machine learning and physical models for geosciences. *Nat Rev Earth Environ* **4**, 552–567 (2023).
65. Bowman, K., Cressie, N., Qu, X. & Hall, A. A Hierarchical Statistical Framework for Emergent Constraints: Application to Snow-Albedo Feedback. *Geophysical Research Letters* **45**, (2018).
66. Kochkov, D. *et al.* Neural general circulation models for weather and climate. *Nature* (2024) doi:10.1038/s41586-024-07744-y.
67. Schneider, T., Lan, S., Stuart, A. & Teixeira, J. Earth System Modeling 2.0: A Blueprint for Models That Learn From Observations and Targeted High-Resolution Simulations. *Geophysical Research Letters* **44**, 12,396-12,417 (2017).
68. Fang, J. & Gentine, P. Exploring Optimal Complexity for Water Stress Representation in Terrestrial Carbon Models: A Hybrid-Machine Learning Model Approach. *Authorea Preprints* (2024).
69. Glorot, X. & Bengio, Y. Understanding the difficulty of training deep feedforward neural networks. in *Proceedings of the Thirteenth International Conference on Artificial Intelligence and Statistics* 249–256 (JMLR Workshop and Conference Proceedings, 2010).

70. Bloom, A. A. & Williams, M. Constraining ecosystem carbon dynamics in a data-limited world: Integrating ecological 'common sense' in a model-data fusion framework. *Biogeosciences* **12**, 1299–1315 (2015).
71. Williams, M., Schwarz, P. A., Law, B. E., Irvine, J. & Kurpius, M. R. An improved analysis of forest carbon dynamics using data assimilation. *Global Change Biology* **11**, 89–105 (2005).
72. Zhou, S., Yu, B., Huang, Y. & Wang, G. Daily underlying water use efficiency for AmeriFlux sites. *Journal of Geophysical Research: Biogeosciences* **120**, 887–902 (2015).
73. Boese, S., Jung, M., Carvalhais, N., Teuling, A. J. & Reichstein, M. Carbon-water flux coupling under progressive drought. *Biogeosciences* **16**, 2557–2572 (2019).
74. Levine, P. A. *et al.* Water Stress Dominates 21st-Century Tropical Land Carbon Uptake. *Global Biogeochemical Cycles* **37**, e2023GB007702 (2023).
75. Worden, M. A. *et al.* Inferred drought-induced plant allocation shifts and their impact on drought legacy at a tropical forest site. *Global Change Biology* **30**, e17287 (2024).
76. Yang, Y. *et al.* CARDAMOM-FluxVal version 1.0: a FLUXNET-based validation system for CARDAMOM carbon and water flux estimates. *Geosci. Model Dev.* **15**, 1789–1802 (2022).
77. Williams, M. Global Carbon Cycle Data Assimilation Using Earth Observation: The CARDAMOM Approach. in *Land Carbon Cycle Modeling* 225–235 (CRC Press, 2022).
78. Quetin, G. R., Bloom, A. A., Bowman, K. W. & Konings, A. G. Carbon Flux Variability From a Relatively Simple Ecosystem Model With Assimilated Data Is Consistent With Terrestrial Biosphere Model Estimates. *J Adv Model Earth Syst* **12**, (2020).
79. Sun, Y. *et al.* From remotely-sensed solar-induced chlorophyll fluorescence to ecosystem structure, function, and service: Part II—Harnessing data. *Global Change Biology* **29**, 2893–2925 (2023).
80. Bradbury, J. *et al.* JAX: composable transformations of Python+ NumPy programs. (2018).
81. Friedl, M. & Sulla-Menashe, D. MODIS/Terra+Aqua Land Cover Type Yearly L3 Global 0.05Deg CMG V061. https://doi.org/10.5067/MODIS/MCD12C1.061 (2022).
82. Hersbach, H. *et al.* ERA5 hourly data on single levels from 1940 to present. Copernicus Climate Change Service (C3S) Climate Data Store (CDS) (2023).
83. USGS EROS Data Center. Global 30 Arc-Second Elevation (GTOPO30). https://doi.org/10.5066/F7DF6PQS (1996).
84. Gesch, D. B., Verdin, K. L. & Greenlee, S. K. New land surface digital elevation model covers the Earth. *Eos, Transactions American Geophysical Union* **80**, 69–70 (1999).
85. Besnard, S. *et al.* Global 1km forest age datasets. BGI Data Portal https://doi.org/10.17871/ForestAgeBGI.2021 (2021).
86. Besnard, S. *et al.* Mapping global forest age from forest inventories, biomass and climate data. *Earth System Science Data* **13**, 4881–4896 (2021).
87. Lang, N., Jetz, W., Schindler, K. & Wegner, J. D. A high-resolution canopy height model of the Earth. *Nat Ecol Evol* **7**, 1778–1789 (2023).
88. Wieder, W. R., Boehnert, J., Bonan, G. B. & Langseth, M. Regridded Harmonized World Soil Database v1.2. Oak Ridge National Laboratory Distributed Active Archive Center https://doi.org/10.3334/ORNLDAAC/1247 (2014).
89. Lan, X., Tans, P. & Thoning, K. W. Trends in globally-averaged CO2 determined from NOAA Global Monitoring Laboratory measurements. https://doi.org/10.15138/9N0H-ZH07 (2024).
90. Chen, Y. *et al.* Multi-decadal trends and variability in burned area from the fifth version of the Global Fire Emissions Database (GFED5). *Earth System Science Data* **15**, 5227–5259 (2023).
91. Fang, J. *et al.* Long-term Continuous SIF-informed Photosynthesis Proxy reconstructed with MODIS surface reflectance (LCSPP-MODIS), 2001-2023. Zenodo https://doi.org/10.5281/zenodo.11658088 (2025).
92. Fang, J. *et al.* Reconstruction of a long-term spatially contiguous solar-induced fluorescence (LCSIF) over 1982-2022. *arXiv preprint arXiv:2311.14987* (2023).
93. Lin, W. *et al.* Reprocessed MODIS Version 6.1 Leaf Area Index dataset. 4TU.ResearchData https://doi.org/10.4121/21858717.v2 (2023).

94. Lin, W. *et al.* Reprocessed MODIS Version 6.1 Leaf Area Index Dataset and Its Evaluation for Land Surface and Climate Modeling. *Remote Sensing* **15**, 1780 (2023).
95. Xu, L. *et al.* Changes in global terrestrial live biomass over the 21st century. *Sci. Adv.* **7**, eabe9829 (2021).
96. Jung, M. *et al.* Scaling carbon fluxes from eddy covariance sites to globe: synthesis and evaluation of the FLUXCOM approach. *Biogeosciences* **17**, 1343–1365 (2020).
97. Fan, N. *et al.* Global apparent temperature sensitivity of terrestrial carbon turnover modulated by hydrometeorological factors. *Nat. Geosci.* **15**, 989–994 (2022).
98. Liu, J. *et al.* Carbon Monitoring System Flux Net Biosphere Exchange 2020 (CMS-Flux NBE 2020). *Earth System Science Data* **13**, 299–330 (2021).
99. Liu, J. *et al.* Carbon monitoring system flux estimation and attribution: impact of ACOS-GOSAT $XCO_2$ sampling on the inference of terrestrial biospheric sources and sinks. *Tellus B: Chemical and Physical Meteorology* **66**, 22486 (2014).
100. Liu, J. *et al.* Contrasting carbon cycle responses of the tropical continents to the 2015–2016 El Niño. *Science* **358**, eaam5690 (2017).
101. Pastorello, G. *et al.* The FLUXNET2015 dataset and the ONEFlux processing pipeline for eddy covariance data. *Scientific data* **7**, 225 (2020).
102. Isaac, P. *et al.* OzFlux data: network integration from collection to curation. *Biogeosciences* **14**, 2903–2928 (2017).
103. Reichstein, M. *et al.* On the separation of net ecosystem exchange into assimilation and ecosystem respiration: review and improved algorithm. *Global Change Biology* **11**, 1424–1439 (2005).
104. Chevan, A. & Sutherland, M. Hierarchical Partitioning. (2024).
105. Lundberg, S. M. & Lee, S.-I. A Unified Approach to Interpreting Model Predictions. in *Advances in Neural Information Processing Systems* vol. 30 (Curran Associates, Inc., 2017